%% file: d1882f.tex
\documentstyle[twoside]{article}

\catcode`\@=11
\long\def\@makefntext#1{
\protect\noindent \hbox to 3.2pt {\hskip-.9pt  
$^{{\eightrm\@thefnmark}}$\hfil}#1\hfill}		

\def\@makefnmark{\hbox to 0pt{$^{\@thefnmark}$\hss}}	
	
\def\ps@myheadings{\let\@mkboth\@gobbletwo
\def\@oddhead{\hbox{}
\rightmark\hfil\eightrm\thepage}   
\def\@oddfoot{}\def\@evenhead{\eightrm\thepage\hfil
\leftmark\hbox{}}\def\@evenfoot{}
\def\sectionmark##1{}\def\subsectionmark##1{}}

\oddsidemargin=\evensidemargin
\newcounter{sectionc}\newcounter{subsectionc}\newcounter{subsubsectionc}
\renewcommand{\section}[1] {\vspace{12pt}\addtocounter{sectionc}{1} 
\setcounter{subsectionc}{0}\setcounter{subsubsectionc}{0}\noindent 
	{\tenbf\thesectionc. #1}\par\vspace{5pt}}
\renewcommand{\subsection}[1] {\vspace{12pt}\addtocounter{subsectionc}{1} 
	\setcounter{subsubsectionc}{0}\noindent 
	{\bf\thesectionc.\thesubsectionc. {\kern1pt \bfit #1}}\par\vspace{5pt}}
\renewcommand{\subsubsection}[1] {\vspace{12pt}\addtocounter{subsubsectionc}{1}
	\noindent{\tenrm\thesectionc.\thesubsectionc.\thesubsubsectionc.
	{\kern1pt \tenit #1}}\par\vspace{5pt}}

\newcommand{\textlineskip}{\baselineskip=13pt}
\newcommand{\smalllineskip}{\baselineskip=10pt}

\def\eightcirc{
\begin{picture}(0,0)
\put(4.4,1.8){\circle{6.5}}
\end{picture}}
\def\eightcopyright{\eightcirc\kern2.7pt\hbox{\eightrm c}} 

\newcommand{\copyrightheading}[1]
	{\vspace*{-2.5cm}\smalllineskip{\flushleft
        {\footnotesize 
        Comptes Rendus Acad. Sci. 94 (1882) 1456-1459
        }\\
	 }}

\newcounter{itemlistc}
\newcounter{romanlistc}
\newcounter{alphlistc}
\newcounter{arabiclistc}

\def\@citex[#1]#2{\if@filesw\immediate\write\@auxout
	{\string\citation{#2}}\fi
\def\@citea{}\@cite{\@for\@citeb:=#2\do
	{\@citea\def\@citea{,}\@ifundefined
	{b@\@citeb}{{\bf ?}\@warning
	{Citation `\@citeb' on page \thepage \space undefined}}
	{\csname b@\@citeb\endcsname}}}{#1}}

\newif\if@cghi
\def\cite{\@cghitrue\@ifnextchar [{\@tempswatrue
	\@citex}{\@tempswafalse\@citex[]}}
\def\citelow{\@cghifalse\@ifnextchar [{\@tempswatrue
	\@citex}{\@tempswafalse\@citex[]}}
\def\@cite#1#2{{$\null^{#1}$\if@tempswa\typeout
	{IJCGA warning: optional citation argument 
	ignored: `#2'} \fi}}

\def\@refcitex[#1]#2{\if@filesw\immediate\write\@auxout
	{\string\citation{#2}}\fi
\def\@citea{}\@refcite{\@for\@citeb:=#2\do
	{\@citea\def\@citea{, }\@ifundefined
	{b@\@citeb}{{\bf ?}\@warning
	{Citation `\@citeb' on page \thepage \space undefined}}
	\hbox{\csname b@\@citeb\endcsname}}}{#1}}

\def\@refcite#1#2{{#1\if@tempswa\typeout
        {IJCGA warning: optional citation argument
	ignored: `#2'} \fi}}

\def\refcite{\@ifnextchar[{\@tempswatrue
	\@refcitex}{\@tempswafalse\@refcitex[]}}

\def\pmb#1{\setbox0=\hbox{#1}
	\kern-.025em\copy0\kern-\wd0
	\kern.05em\copy0\kern-\wd0
	\kern-.025em\raise.0433em\box0}

\def\fnt#1#2{\footnotetext{\kern-.3em
	{$^{\mbox{\scriptsize #1}}$}{#2}}}

\font\tenrm=cmr10
\font\tenit=cmti10 
\font\tenbf=cmbx10
\font\bfit=cmbxti10 at 10pt
\font\ninerm=cmr9

\font\eightrm=cmr8

\def\qed{\hbox{${\vcenter{\vbox{			
   \hrule height 0.4pt\hbox{\vrule width 0.4pt height 6pt
   \kern5pt\vrule width 0.4pt}\hrule height 0.4pt}}}$}}

\begin{document}

\input gdar.tex

\newpage



\normalsize\textlineskip
\thispagestyle{empty}

\copyrightheading{}			

\vspace*{0.88truein}

\centerline{ANALYSE MATH\'EMATIQUE.}
\bigskip
\centerline{\bf SUR UNE PROPOSITION RELATIVE AUX \'EQUATIONS LIN\'EAIRES}
\vspace*{0.035truein}
\vspace*{0.37truein}
\centerline{\footnotesize Note de {\cal M}. G. Darboux}
\vspace*{0.015truein}
\centerline{[\footnotesize{\it LaTex par} {\cal M}. H.C. Rosu (1999)]}
\baselineskip=10pt
\vspace*{10pt}
\vspace*{0.225truein}

\vspace*{0.21truein}


\textlineskip                  
\vspace*{12pt}                 

\vspace*{1pt}\textlineskip	
\vspace*{-0.5pt}
\noindent


\noindent




{\bf 1}.
Dans ma derni\`ere Communication, j'ai montr\'e comment, toutes les fois
que l'on saura int\'egrer
l'\'equation
$$
\frac{d^2y}{dx^2}= y[f(x)+m]~.
$$
pour toutes les valeurs de la constante $m$, on pourra obtenir une suite
illimit\'ee d'equations, contenant de la m\^eme mani\`ere un param\`etre
variable,
et dont l'int\'egration sera possible pour toutes les valeurs du param\`etre.

Voici la d\'emonstration qui m'a conduit \`a ce r\'esultat:

\bigskip

{\bf 2}. Soit, d'une mani\`ere g\'en\'erale,
\begin{equation}
y ^{''}+Py^{'}+Qy=0
\label{1}
\end{equation}
une \'equation lin\'eaire du second ordre. $y$ d\'esignant une
int\'egrale quelconque de cette \'equation, je consid\`ere l'\'equation
lin\'eaire qui admet pour int\'egrale
\begin{equation} \label{2}
u=Ay+By^{'}~,
\end{equation}
$A$ et $B$ \'etant des fonctions quelconques de $x$, et je vais d'abord
chercher quelle relation il faut \'etablir entre $A$ et $B$ pour que cette
\'equation soit de la forme
\begin{equation} \label{3}
u^{''}+Pu^{'}+Q_{1}u=0~,
\end{equation}
le coefficient de $u^{'}$ \'etant le m\^eme que celui de $y^{'}$ dans la
premi\`ere \'equation.
Pour cela il faut \'evidemment que, si
$u_1$, $u_2$ sont deux int\'egrales particuli\`eres correspondantes
\`a deux int\'egrales $y_1$, $y_2$ de l'\'equation (1), on ait
$$
u_{1}u_{2}^{'}-u_{2}u_{1}^{'}=C(y_1y_{2}^{'}-y_{2}y_{1}^{'})~,
$$
$C$ d\'esignant une constante quelconque, que l'on peut \'evidemment remplacer
par l'unit\'e. On est ainsi conduit \`a la relation diff\'erentielle
$$
A^{2}+AB^{'}-BA^{'}-ABP+B^{2}Q=1~,
$$
qui doit avoir lieu entre $A$ et $B$.

\bigskip

{\bf 3}.
La forme de cette relation permet de trouver les fonctions les plus
g\'en\'erales qui peuvent y satisfaire. Prenons pour inconnue auxiliaire
le rapport de $A$ \`a $B$, et posons
$$
A=-\lambda B~.
$$
Nous aurons
$$
B^2=\frac{1}{\lambda ^2 +\lambda ^{'}+\lambda P +Q}~.
$$
Pour plus de sym\'etrie rempla\c cons $\lambda$ pour
$\frac{\theta ^{'}}{\theta}$,
$\theta$ d\'esignant un autre fonction, l'expression de $u$ deviendra
$$
u=\frac{\theta y^{'}-y\theta ^{'}}{\sqrt{\theta(\theta ^{''}+P\theta ^{'}
+Q\theta)}}~,
$$
ou plus simplement
\begin{equation} \label{18}
u=\frac{\theta y^{'}-y\theta ^{'}}{H}~,
\end{equation}
en posant
\begin{equation}  \label{19}
H={\sqrt{\theta(\theta ^{''}+P\theta ^{'}
+Q\theta})}~.
\end{equation}

\bigskip

{\bf 4}.
Cela pos\'e, il est tr\`es ais\'e de former l'\'equation \`a laquelle
satisfait la valeur du $u$, et l'on trouve ainsi
\begin{equation}  \label{20}
u^{''}+Pu^{'}+u\Big[\frac{H^2}{\theta ^{2}}-H\frac{d^2\frac{1}{H}}{dx^2}+
H\frac{d}{dx}\left(\frac{P}{H}\right)\Big]=0~.
\end{equation}
On peut faire diverses applications de cette \'equation.

\bigskip

{\bf 5}.
Supposons, par exemple, que l'on sache int\'egrer, quelle que soit la
constante $m$, l'\'equation
\begin{equation}  \label{21}
y^{''}+Py^{'}+(Q-mR)y=0~.
\end{equation}
On aura ici
$$
u=\frac{y\theta ^{'}-\theta y^{'}}{\sqrt{\theta (\theta ^{''}+P\theta ^{'}
+Q\theta -mR\theta)}}~.
$$
Prenons pour $\theta$ une int\'egrale de l'\'equation
\begin{equation}  \label{8}
\theta ^{''}+P\theta ^{'}+Q\theta =0~;
\end{equation}
$u$ se r\'eduira, \`a une constante pr\'es, \`a la fonction
\begin{equation} \label{9}
u=\frac{y^{'}-\frac{\theta ^{'}}{\theta}y}{\sqrt{R}}~,
\end{equation}
et cette fonction satisfera \`a l'\'equation
\begin{equation}  \label{10}
u^{''}+Pu^{'}+u\Big[-mR-\theta \sqrt{R}\frac{d^2}{dx^2}
\left(\frac{1}{\theta \sqrt{R}}\right)
+\theta \sqrt{R}\frac{d}{dx}\left(\frac{P}
{\theta \sqrt{R}}\right)\Big]=0~,
\end{equation}
qui contient le param\`etre $m$ de la m\^eme mani\`ere que la pr\'ec\'edente.

\bigskip

{\bf 6}.
En particulier, si l'on consid\`ere l'\'equation
\begin{equation}  \label{11}
y^{''}=y[f(x)+m]~,
\end{equation}
on voit que, $\theta$ satisfaisant \`a l'\'equation
$$
\theta ^{''}=f(x)\theta
$$
la fonction
$$
u=y^{'}-\frac{\theta ^{'}}{\theta}y
$$
sera une int\'egrale de l'\'equation
\begin{equation}  \label{12}
u^{''}=u\Big[m+\theta \frac{d^2\frac{1}{\theta}}{dx^2}\Big]
\end{equation}
toutes les fois que $y$ satisfera \`a l'\'equation (11).

\bigskip

{\bf 7}.
On pourrait craindre que le proc\'ed\'e que nous venons d'indiquer ne
conduise qu'\`a un nombre limit\'e d'\'equations r\'eellement distinctes.
Mais il suffit de prendre un exemple num\'erique pour se convaincre que
l'on pourra obtenir une suite ind\'efinie d\'equations diff\'erentes.
Consid\'erons, par exemple, l'\'equation
$$
y^{''}=my~.
$$
En employant la solution $\theta =x$, on aura l'\'equation
$$
y^{''}=\Big[\frac{1\cdot 2}{x^2}+m\Big]y~.
$$
Appliquant la m\^eme methode \`a cette \'equation, en prenant maintenant
$\theta =x^2$, on aura
$$
y^{''}=\Big[\frac{2\cdot 3}{x^2}+m\Big]y~,
$$
et ainsi de suite.

\bigskip

{\bf 8}.
Le raisonnements pr\'ec\'edents reposent sur ce fait que le
coefficient de $u$ dans l'\'equation (6) se compose de deux parties, l'une
du degr\'e z\'ero, l'autre du second degr\'e par rapport \`a $H$.
On peut utiliser autrement cette propri\'et\'e de l'\'equation (6), et
obtenir plusieurs th\'eor\'`emes analogues au pr\'ec\'edent.
Par exemple, on peut ramener l'\'equation
$$
y^{''}=mf(x)y
$$
\`a une autre qui sera de la forme
$$
y^{''}=[m\phi (x)+\psi (x)]y~,
$$
et cela de plusieurs mani\`eres diff\'erentes.

\newpage

\input denin.tex

\newpage

\input d1882rin.tex

\newpage

\input d1882sin.tex

%% file: gdar.tex

\pagestyle{empty}

\begin{center}
{\it Los Alamos Electronic Archives: physics/9908003}
\end{center}

\bigskip

\begin{center}
{\bf MAKING OLD SEMINAL RESULTS WORLD-WIDE AVAILABLE !}
\end{center}

$\;$\\
$\;$\\
$\;$\\
$\;$\\
$\;$\\
$\;$\\

\begin{center}

{\bf FORWARD}

\end{center}

\bigskip

\noindent
The short {\em Comptes Rendus} Note of Darboux, first published in 1882, and
also included in his course ``Th\'eorie des Surfaces", second volume,
page 210 (1889),
was not recognized as important for about one century. In his famous
textbook of 1926, Ince
mentioned particular cases of Darboux's result as exercises.
However, Ince's formulation is very close to the framework of
supersymmetric quantum mechanics, and therefore should be considered as a
valuable contribution. In 1979, V.B. Matveev realized the importance of
the result and introduced the terminology ``Darboux transformations", which
he emphasized once again in the title of his 1991
beautiful book co-authored with M.A. Salle.
During the eighties, many SUSY QM authors became aware of this old seminal
result, that registered in this way a worthy citation record.

\bigskip

\noindent
Here, for the benefit of the active authors and other
interested people,
I offer the original French text of Darboux's Note, together with
English, Romanian and Spanish translations. This instructive Note has 8
points displaying 12 numbered and 14
unnumbered formulae, many of them particular cases of the others. The result,
as masterly presented by Darboux, is by far more general than SUSY QM,
which is merely the particular case $P=0$, $B=\pm 1$, $R=1$, in the 
notations of Darboux. In the published text, I have detected and corrected
three misprints.
Two of them are in the unnumbered formula following Eq. 7, where, after
the $R$ symbol, I added a $\theta$ and a right parenthesis.
The other
misprint is in Eq. 10, where there was a missing right parenthesis
closing the second derivative in the coefficient of $u$.

\bigskip
\bigskip

\hfill ${\cal H}$ ${\cal C}$ ${\cal R}$

\bigskip

\hfill 8. 2. 1999


%% file: denin.tex


\centerline{\footnotesize Comptes Rendus Acad. Sci. 94 (1882) 1456-1459}

\bigskip

\centerline{MATHEMATICAL ANALYSIS.}
\bigskip
\centerline{\bf ON A PROPOSITION RELATIVE TO LINEAR EQUATIONS}
\vspace*{0.035truein}
\vspace*{0.37truein}
\centerline{\footnotesize Note by Mr. G. Darboux}
\vspace*{0.015truein}
\centerline{[\footnotesize{\it translated from French by Mr. H.C. Rosu} (1999)]}
\baselineskip=10pt
\vspace*{10pt}
\vspace*{0.225truein}

\vspace*{0.21truein}


\textlineskip                  
\vspace*{12pt}                 

\vspace*{1pt}\textlineskip	
\vspace*{-0.5pt}
\noindent


\noindent



\setcounter{equation}{0}

{\bf 1}.
In my previous Communication, I have shown that whenever one
knows to integrate the equation
$$
\frac{d^2y}{dx^2}= y[f(x)+m]~.
$$
for all the values of the constant $m$, one can obtain an infinite set of
equations, displaying the variable parameter in the same way,
which are integrable for any value of the parameter.

Here, I present the proof that guided me toward getting this result:

\bigskip

{\bf 2}. Let
\begin{equation}
y ^{''}+Py^{'}+Qy=0
\label{1}
\end{equation}
be a general linear equation of second order. If $y$ is whatever
integral of this equation, I consider the linear equation with the
following integrals
\begin{equation} \label{2}
u=Ay+By^{'}~,
\end{equation}
where $A$ and $B$ are two arbitrary functions of $x$. First, I will look for
the required relationship between $A$ and $B$ to have (2) of the form
\begin{equation} \label{3}
u^{''}+Pu^{'}+Q_{1}u=0~,
\end{equation}
where the coefficient of $u^{'}$ is the same as of $y^{'}$ in (1).
It is obvious that one should comply with the following condition
$$
u_{1}u_{2}^{'}-u_{2}u_{1}^{'}=C(y_1y_{2}^{'}-y_{2}y_{1}^{'})~,
$$
for any two particular integrals $u_1$ and $u_2$ of (2) corresponding to
two particular integrals $y_1$ and $y_2$ of (1). $C$ is an arbitrary
constant that we can obviously put equal to unity. This will lead us to
the differential relationship
$$
A^{2}+AB^{'}-BA^{'}-ABP+B^{2}Q=1~,
$$
involving $A$, $B$, $P$, $Q$.

\bigskip

{\bf 3}.
The form of this relationship allows to find the most general
functions fulfilling it.
We employ the quotient of $A$ and $B$ as an auxiliary unknown, and write
$$
A=-\lambda B~.
$$
We shall have
$$
B^2=\frac{1}{\lambda ^2 +\lambda ^{'}+\lambda P +Q}~.
$$
To achieve more symmetry, we write $\lambda =\frac{\theta ^{'}}{\theta}$,
where $\theta$ is another function. Then, the expression for $u$ will be
$$
u=\frac{\theta y^{'}-y\theta ^{'}}{\sqrt{\theta(\theta ^{''}+P\theta ^{'}
+Q\theta)}}~,
$$
or, in a shorter notation
\begin{equation} \label{18}
u=\frac{\theta y^{'}-y\theta ^{'}}{H}~,
\end{equation}
where
\begin{equation}  \label{19}
H={\sqrt{\theta(\theta ^{''}+P\theta ^{'}
+Q\theta})}~.
\end{equation}

\bigskip

{\bf 4}.
Once we have determined this, it is quite easy to get the equation for $u$,
which is
\begin{equation}  \label{20}
u^{''}+Pu^{'}+u\Big[\frac{H^2}{\theta ^{2}}-H\frac{d^2\frac{1}{H}}{dx^2}+
H\frac{d}{dx}\left(\frac{P}{H}\right)\Big]=0~.
\end{equation}
One can make various applications of this equation.

\bigskip

{\bf 5}.
Suppose, for example, that we know how to integrate, for any value of the
constant $m$, the following equation
\begin{equation}  \label{21}
y^{''}+Py^{'}+(Q-mR)y=0~.
\end{equation}
Here, we shall have
$$
u=\frac{y\theta ^{'}-\theta y^{'}}{\sqrt{\theta (\theta ^{''}+P\theta ^{'}
+Q\theta -mR\theta)}}~.
$$
Let us take for $\theta$ an integral of the equation
\begin{equation}  \label{8}
\theta ^{''}+P\theta ^{'}+Q\theta =0~;
\end{equation}
up to a constant, $u$ will be the function
\begin{equation} \label{9}
u=\frac{y^{'}-\frac{\theta ^{'}}{\theta}y}{\sqrt{R}}~,
\end{equation}
satisfying the equation
\begin{equation}  \label{10}
u^{''}+Pu^{'}+u\Big[-mR-\theta \sqrt{R}\frac{d^2}{dx^2}
\left(\frac{1}{\theta \sqrt{R}}\right)+\theta \sqrt{R}\frac{d}{dx}
\left(\frac{P}{\theta \sqrt{R}}\right)\Big]=0~,
\end{equation}
where the $m$ parameter appears in the same manner as in
the previous equation.

\bigskip

{\bf 6}.
In particular, for the equation
\begin{equation}  \label{11}
y^{''}=y[f(x)+m]~,
\end{equation}
one can see that if $\theta$ satisfies the equation
$$
\theta ^{''}=f(x)\theta
$$
the function
$$
u=y^{'}-\frac{\theta ^{'}}{\theta}y
$$
will be an integral of the equation
\begin{equation}  \label{12}
u^{''}=u\Big[m+\theta \frac{d^2\frac{1}{\theta}}{dx^2}\Big]
\end{equation}
whenever $y$ satisfies the equation (11).

\bigskip

{\bf 7}.
One can fear that our procedure will only lead to a finite number of really
different equations. However, it is sufficient to take a numerical
example in order to convince ourselves that an infinite set of
different equations can be obtained. Let the example be the equation
$$
y^{''}=my~.
$$
Employing the solution $\theta =x$, we shall get
$$
y^{''}=\Big[\frac{1\cdot 2}{x^2}+m\Big]y~.
$$
Applying the same method to the latter equation, but
taking now $\theta =x^2$, we shall have
$$
y^{''}=\Big[\frac{2\cdot 3}{x^2}+m\Big]y~,
$$
and so on.

\bigskip

{\bf 8}.
The previous arguments rely on the fact that the coefficient of $u$ in (6)
entails two terms, one of zero degree and the other of second degree with
respect to $H$. This property of (6) can be used in a different way in order
to get several analog theorems to that presented here.
For example, we can bring the equation
$$
y^{''}=mf(x)y
$$
to another one of the form
$$
y^{''}=[m\phi (x)+\psi (x)]y~,
$$
in several different ways.

%% file: d1882rin.tex


\centerline{\footnotesize Comptes Rendus Acad. Sci. 94 (1882) 1456-1459}

\bigskip

\centerline{ANALIZ\v A MATEMATIC\v A.}
\bigskip
\centerline{\bf ASUPRA UNEI PROPOZI\c TII RELATIVE LA ECUA\c TIILE LINEARE}
\vspace*{0.035truein}
\vspace*{0.37truein}
\centerline{\footnotesize Not\v a a {\it Dlui}. G. Darboux}
\vspace*{0.015truein}
\centerline{[\footnotesize{\it pus\v a \^{\i}n  LaTex \c si tradus\v a
de c\v atre Dl.}
H.C. Rosu (1999)]}
\baselineskip=10pt
\vspace*{10pt}
\vspace*{0.225truein}

\vspace*{0.21truein}


\textlineskip                  
\vspace*{12pt}                 

\vspace*{1pt}\textlineskip	
\vspace*{-0.5pt}
\noindent


\noindent



\setcounter{equation}{0}

{\bf 1}.
In ultima mea Comunicare, am ar\v atat cum in fiecare caz in care se \c stie
cum se integreaz\v a ecua\c tia
$$
\frac{d^2y}{dx^2}= y[f(x)+m]~,
$$
pentru toate valorile constantei $m$, se va putea ob\c tine un \c sir
infinit de ecua\c tii, con\c tin\^{\i}nd in acela\c si mod un parametru
variabil,
\c si a c\v aror integrare este posibil\v a pentru toate valorile
parametrului.

Iat\v a aici demonstra\c tia care m-a condus la acest rezultat:

\bigskip

{\bf 2}. Fie, \^{\i}n general,
\begin{equation}
y ^{''}+Py^{'}+Qy=0
\label{1}
\end{equation}
o ecua\c tie linear\v a de ordinul doi. $y$ fiind o
integral\v a oarecare a acestei ecua\c tii, voi considera ecua\c tia
linear\v a care admite ca integral\v a
\begin{equation} \label{2}
u=Ay+By^{'}~,
\end{equation}
$A$ \c si $B$ fiind func\c tii oarecare de $x$, \c si voi c\v auta
mai \^{\i}nt\^{\i}i care este rela\c tia \^{\i}ntre $A$ \c si $B$ pentru ca
aceast\v a ecua\c tie sa fie de forma
\begin{equation} \label{3}
u^{''}+Pu^{'}+Q_{1}u=0~,
\end{equation}
coeficientul lui $u^{'}$ fiind identic cu cel al lui $y^{'}$ din prima
ecua\c tie.
Pentru aceasta, dac\v a
$u_1$, $u_2$ sunt dou\v a integrale particulare ale lui (3)
corespunz\v atoare la
dou\v a integrale particulare $y_1$, $y_2$ ale ecua\c tiei (1), trebuie
evident s\v a avem
$$
u_{1}u_{2}^{'}-u_{2}u_{1}^{'}=C(y_1y_{2}^{'}-y_{2}y_{1}^{'})~,
$$
$C$ fiind o constant\v a oarecare, ce se poate substitui cu unitatea.
Astfel, suntem condu\c si la rela\c tia diferen\c tial\v a
$$
A^{2}+AB^{'}-BA^{'}-ABP+B^{2}Q=1~,
$$
care trebuie s\v a aib\v a loc \^{\i}ntre $A$ \c si $B$.

\bigskip

{\bf 3}.
Forma acestei rela\c tii permite ob\c tinerea celor mai generale
func\c tii care o pot satisface. Pentru aceasta,
s\v a lu\v am ca necunoscut\v a auxiliar\v a
raportul \^{\i}ntre $A$ \c si $B$, \c si s\v a punem
$$
A=-\lambda B~.
$$
Vom avea
$$
B^2=\frac{1}{\lambda ^2 +\lambda ^{'}+\lambda P +Q}~.
$$
Pentru mai mult\v a simetrie s\v a \^{\i}nlocuim pe $\lambda$ cu
$\frac{\theta ^{'}}{\theta}$,
$\theta$ fiind o alt\v a func\c tie; expresia lui $u$ devine
$$
u=\frac{\theta y^{'}-y\theta ^{'}}{\sqrt{\theta(\theta ^{''}+P\theta ^{'}
+Q\theta)}}~,
$$
sau mai simplu
\begin{equation} \label{18}
u=\frac{\theta y^{'}-y\theta ^{'}}{H}~,
\end{equation}
unde am pus
\begin{equation}  \label{19}
H={\sqrt{\theta(\theta ^{''}+P\theta ^{'}
+Q\theta})}~.
\end{equation}

\bigskip

{\bf 4}.
Toate acestea stabilite, este foarte u\c sor de a forma ecua\c tia
pe care o satisface $u$, pentru care am g\v asit
\begin{equation}  \label{20}
u^{''}+Pu^{'}+u\Big[\frac{H^2}{\theta ^{2}}-H\frac{d^2\frac{1}{H}}{dx^2}+
H\frac{d}{dx}\left(\frac{P}{H}\right)\Big]=0~.
\end{equation}
Se pot face diverse aplica\c tii ale acestei ecua\c tii.

\bigskip

{\bf 5}.
S\v a presupunem, de exemplu, c\v a oricare ar fi
constanta $m$, \c stim sa integr\v am ecua\c tia
\begin{equation}  \label{21}
y^{''}+Py^{'}+(Q-mR)y=0~.
\end{equation}
Vom avea \^{\i}n acest caz
$$
u=\frac{y\theta ^{'}-\theta y^{'}}{\sqrt{\theta (\theta ^{''}+P\theta ^{'}
+Q\theta -mR\theta)}}~.
$$
Ca func\c tie $\theta$ vom lua o integral\v a a ecua\c tiei
\begin{equation}  \label{8}
\theta ^{''}+P\theta ^{'}+Q\theta =0~;
\end{equation}
\c si p\^{\i}n\v a la o constant\v a arbitrar\v a, $u$ se va reduce la
func\c tia
\begin{equation} \label{9}
u=\frac{y^{'}-\frac{\theta ^{'}}{\theta}y}{\sqrt{R}}~,
\end{equation}
care va satisface ecua\c tia
\begin{equation}  \label{10}
u^{''}+Pu^{'}+u\Big[-mR-\theta \sqrt{R}\frac{d^2}{dx^2}
\left(\frac{1}{\theta \sqrt{R}}\right)
+\theta \sqrt{R}\frac{d}{dx}\left(\frac{P}
{\theta \sqrt{R}}\right)\Big]=0~,
\end{equation}
unde parametrul $m$ se prezint\v a la fel ca \^{\i}n precedenta.

\bigskip

{\bf 6}.
In particular, dac\v a se consider\v a ecua\c tia
\begin{equation}  \label{11}
y^{''}=y[f(x)+m]~,
\end{equation}
\^{\i}n cazul \^{\i}n care $\theta$ satisface ecua\c tia
$$
\theta ^{''}=f(x)\theta
$$
se vede c\v a func\c tia
$$
u=y^{'}-\frac{\theta ^{'}}{\theta}y
$$
va fi o  integral\v a a ecua\c tiei
\begin{equation}  \label{12}
u^{''}=u\Big[m+\theta \frac{d^2\frac{1}{\theta}}{dx^2}\Big]
\end{equation}
ori de c\^{\i}te ori $y$ satisface ecua\c tia (11).

\bigskip

{\bf 7}.
Ne-am putea teme c\v a procedura pe care am indicat-o nu
conduce dec\^{\i}t la un num\v ar finit de ecua\c tii cu adev\v arat distincte.
Dar, este suficient a da un exemplu numeric
pentru a ne convinge c\v a vom putea ob\c tine un \c sir infinit de
ecua\c tii diferite.
S\v a consider\v am, ca exemplu, ecua\c tia
$$
y^{''}=my~.
$$
Folosind solu\c tia $\theta =x$, vom avea ecua\c tia
$$
y^{''}=\Big[\frac{1\cdot 2}{x^2}+m\Big]y~.
$$
Aplic\^{\i}nd acestei ecua\c tii aceea\c si metod\v a, lu\^{\i}nd \^{\i}n
continuare $\theta =x^2$, vom avea
$$
y^{''}=\Big[\frac{2\cdot 3}{x^2}+m\Big]y~,
$$
\c si a\c sa mai departe.

\bigskip

{\bf 8}.
Ra\c tionamentele precedente se bazeaz\v a pe faptul c\v a
\^{\i}n ecua\c tia (6) coeficientul lui $u$ se poate descompune \^{\i}n
dou\v a p\v ar\c ti,
una de grad zero \c si alta de grad doi in raport cu $H$.
 Aceasta proprietate a ecua\c tiei
(6) se poate utiliza \c si \^{\i}n alte variante pentru a ob\c tine mai multe
teoreme analoage celei date aici.
De exemplu, se poate aduce ecua\c tia
$$
y^{''}=mf(x)y
$$
la forma
$$
y^{''}=[m\phi (x)+\psi (x)]y~,
$$
\c si aceasta \^{\i}n mai multe moduri diferite.

%% file: d1882sin.tex


\centerline{\footnotesize Comptes Rendus Acad. Sci. 94 (1882) 1456-1459}
\bigskip
\centerline{AN\'ALISIS MATEM\'ATICO.}
\bigskip
\centerline{\bf SOBRE UNA PROPOSICI\'ON RELATIVA A LAS ECUACIONES LINEALES}
\vspace*{0.035truein}
\vspace*{0.37truein}
\centerline{\footnotesize Nota de {\it Sr}. G. Darboux}
\vspace*{0.015truein}
\centerline{[\footnotesize{\it traducci\'on por}
H.C. Rosu (1999)]}
\baselineskip=10pt
\vspace*{10pt}
\vspace*{0.225truein}

\vspace*{0.21truein}


\textlineskip                  
\vspace*{12pt}                 

\vspace*{1pt}\textlineskip	
\vspace*{-0.5pt}
\noindent


\noindent



\setcounter{equation}{0}

{\bf 1}.
En mi \'ultima Comunicaci\'on, he mostrado como, cada vez
que se sabe integrar la ecuaci\'on
$$
\frac{d^2y}{dx^2}= y[f(x)+m]~.
$$
para todos los valores de la constante $m$, se puede obtener una cadena
infinita de ecuaciones, con el par\'ametro $m$ contenido de la misma
forma, y para las cuales la integraci\'on sera posible para todos los
valores del par\'ametro.

Aqu\'{\i} va la demostraci\'on que me ha portado al resultado mencionado:

\bigskip

{\bf 2}. Sea, de una manera general,
\begin{equation}
y ^{''}+Py^{'}+Qy=0
\label{1}
\end{equation}
una ecuaci\'on lineal de segundo orden. $y$ siendo una
integral cualquiera de esta ecuaci\'on, consideremos la ecuaci\'on
lineal que admite como integral
\begin{equation} \label{2}
u=Ay+By^{'}~,
\end{equation}
$A$ y $B$ siendo funciones cualquiera de $x$, y encontraremos primero
la relaci\'on que hay entre $A$ y $B$ para que esta
ecuaci\'on sea de la forma
\begin{equation} \label{3}
u^{''}+Pu^{'}+Q_{1}u=0~,
\end{equation}
el coeficiente de $u^{'}$ siendo el mismo que el de $y^{'}$ en la
primera ecuaci\'on.
Para esto, si
$u_1$, $u_2$ son dos integrales particulares correspondientes
a dos integrales $y_1$, $y_2$ de la ecuaci\'on (1), obviamente se cumple
$$
u_{1}u_{2}^{'}-u_{2}u_{1}^{'}=C(y_1y_{2}^{'}-y_{2}y_{1}^{'})~,
$$
donde $C$ es una constante cualquiera, que puede ser sustituida por la unidad.
De tal manera, se llega a la relaci\'on diferencial
$$
A^{2}+AB^{'}-BA^{'}-ABP+B^{2}Q=1~,
$$
que se requiere entre $A$ y $B$.

\bigskip

{\bf 3}.
La forma que toma esta relaci\'on permite obtener las funciones m\'as
generales que la cumplen. Tomamos como par\'ametro auxiliar
el cociente de $A$ y $B$, y ponemos
$$
A=-\lambda B~.
$$
teniendo
$$
B^2=\frac{1}{\lambda ^2 +\lambda ^{'}+\lambda P +Q}~.
$$
Para lograr m\'as simetria sustituimos $\lambda$ por
$\frac{\theta ^{'}}{\theta}$,
$\theta$ siendo otra funci\'on, la expresi\'on de $u$ ser\'a
$$
u=\frac{\theta y^{'}-y\theta ^{'}}{\sqrt{\theta(\theta ^{''}+P\theta ^{'}
+Q\theta)}}~,
$$
o m\'as sencillo
\begin{equation} \label{18}
u=\frac{\theta y^{'}-y\theta ^{'}}{H}~,
\end{equation}
poniendo
\begin{equation}  \label{19}
H={\sqrt{\theta(\theta ^{''}+P\theta ^{'}
+Q\theta})}~.
\end{equation}

\bigskip

{\bf 4}.
De este modo, se puede facilmente formar la ecuaci\'on a la cual
cumple el valor de $u$, y as\'{\i} se encuentra
\begin{equation}  \label{20}
u^{''}+Pu^{'}+u\Big[\frac{H^2}{\theta ^{2}}-H\frac{d^2\frac{1}{H}}{dx^2}+
H\frac{d}{dx}\left(\frac{P}{H}\right)\Big]=0~.
\end{equation}
Se pueden hacer varias aplicaciones de esta ecuaci\'on.

\bigskip

{\bf 5}.
Suponemos, por ejemplo, que se sabe integrar, para cualquier
constante $m$, la ecuaci\'on
\begin{equation}  \label{21}
y^{''}+Py^{'}+(Q-mR)y=0~.
\end{equation}
En este caso
$$
u=\frac{y\theta ^{'}-\theta y^{'}}{\sqrt{\theta (\theta ^{''}+P\theta ^{'}
+Q\theta -mR\theta)}}~.
$$
Tomamos como $\theta$ una integral de la ecuaci\'on
\begin{equation}  \label{8}
\theta ^{''}+P\theta ^{'}+Q\theta =0~;
\end{equation}
y hasta una constante, $u$ se reduce a la funci\'on
\begin{equation} \label{9}
u=\frac{y^{'}-\frac{\theta ^{'}}{\theta}y}{\sqrt{R}}~,
\end{equation}
que es soluci\'on de la siguiente ecuaci\'on
\begin{equation}  \label{10}
u^{''}+Pu^{'}+u\Big[-mR-\theta \sqrt{R}\frac{d^2}{dx^2}
\left(\frac{1}{\theta \sqrt{R}}\right)
+\theta \sqrt{R}\frac{d}{dx}\left(\frac{P}
{\theta \sqrt{R}}\right)\Big]=0~,
\end{equation}
donde el par\'ametro $m$ ocurre en la misma forma que en la anterior.

\bigskip

{\bf 6}.
En particular, si se considera la ecuaci\'on
\begin{equation}  \label{11}
y^{''}=y[f(x)+m]~,
\end{equation}
se ve que con $\theta$ como soluci\'on de la ecuaci\'on
$$
\theta ^{''}=f(x)\theta~,
$$
la funci\'on
$$
u=y^{'}-\frac{\theta ^{'}}{\theta}y
$$
ser\'a una integral de la ecuaci\'on
\begin{equation}  \label{12}
u^{''}=u\Big[m+\theta \frac{d^2\frac{1}{\theta}}{dx^2}\Big]
\end{equation}
cada vez que $y$ sea soluci\'on de la ecuaci\'on (11).

\bigskip

{\bf 7}.
Uno puede temer que el procedimiento que acabamos de presentar no da
m\'as que un n\'umero finito de ecuaciones de verdad diferentes.
Pero, es suficiente de tomar un ejemplo num\'erico para convencerse que
se podr\'a obtener una cadena infinita de ecuaciones diferentes.
Consideremos, por ejemplo, la ecuaci\'on
$$
y^{''}=my~.
$$
Utilizando la soluci\'on $\theta =x$, se llega a la ecuaci\'on
$$
y^{''}=\Big[\frac{1\cdot 2}{x^2}+m\Big]y~.
$$
Aplicando el mismo m\'etodo a esta \'ultima ecuaci\'on, pero tomando ahora
$\theta =x^2$, se llega a
$$
y^{''}=\Big[\frac{2\cdot 3}{x^2}+m\Big]y~,
$$
y as\'{\i} sucesivamente.

\bigskip

{\bf 8}.
Los argumentos anteriores se basan en el hecho de que el
coeficiente de $u$ en la ecuaci\'on (6) est\'a compuesto de dos partes, una
de grado zero, otra de segundo grado con respecto a $H$.
Se puede utilizar de otra manera esta propiedad de la ecuaci\'on (6), y
obtener m\'as teoremas an\'alogos al anterior.
Por ejemplo, se puede iniciar con la ecuaci\'on
$$
y^{''}=mf(x)y
$$
para llegar a la forma
$$
y^{''}=[m\phi (x)+\psi (x)]y~,
$$
y esto, por varios caminos diferentes.

\end{document}